\documentclass[showpacs,superscriptaddress,twocolumn,pra]{revtex4-1}
\usepackage{color}
\usepackage{amsfonts,amssymb,amsmath}
\usepackage{amsbsy}

\usepackage[vcentermath]{youngtab}

\newcommand{\bra}[1]{\left\langle #1 \right|}
\newcommand{\ket}[1]{\left| #1 \right\rangle}

\usepackage{graphicx}

\begin{document}

\title{Recursive encoding and decoding of the noiseless subsystem for qudits}

\author{Utkan G\"ung\"ord\"u}
\affiliation{Research Center for Quantum Computing, Interdisciplinary Graduate School of Science and Engineering, Kinki University, 3-4-1 Kowakae, Higashi-Osaka, Osaka 577-8502, Japan}
\email{Corresponding author}
\email{utkan@alice.math.kindai.ac.jp}

\author{Chi-Kwong Li}
\affiliation{Department of Mathematics, College of William \& Mary,
Williamsburg, VA 23187-8795, USA. (Year 2011: Department of Mathematics, Hong Kong University
of Science \& Technology, Hong Kong.)}
\email{ckli@math.wm.edu}

\author{Mikio Nakahara}
\affiliation{Research Center for Quantum Computing, Interdisciplinary Graduate School of Science and Engineering, Kinki University, 3-4-1 Kowakae, Higashi-Osaka, Osaka 577-8502, Japan}
\affiliation{Department of Physics, Kinki University, 3-4-1 Kowakae, Higashi-Osaka, Osaka 577-8502, Japan}
\email{nakahara@math.kindai.ac.jp}

\author{Yiu-Tung Poon}
\affiliation{Department of Mathematics, Iowa State University,
Ames, IA 50011, USA.}
\email{ytpoon@iastate.edu}

\author{Nung-Sing Sze}
\affiliation{Department of Applied Mathematics, The Hong Kong Polytechnic University,
Hung Hom, Hong Kong.}
\email{raymond.sze@inet.polyu.edu.hk}


\begin{abstract}
We give a full explanation of the noiseless subsystem that protects a single-qubit against collective errors and the corresponding recursive scheme described by C.-K. Li et. al. [\emph{Phys. Rev. A} \textbf{84}, 044301 (2011)] from a representation theory point of view. Furthermore, we extend the construction to qudits under the influence of collective SU($d$) errors. We find that under this recursive scheme, the asymptotic encoding rate is $1/d$.
\end{abstract}

\maketitle

\section{Introduction}
Quantum computing and quantum information processing make
use of quantum systems as computational resources to
outperform their classical counterparts. It is expected
that a quantum computer solves computationally hard
tasks for a classical computer, such as prime number factorization of a large
number, in a practical time and quantum key distribution
realizes a 100\% secure classical information transmission. 
In spite of this expectation, a working quantum computer
has not become a reality yet. One of the obstacles against
its realization is decoherence. Decoherence is a process
caused by a coupling between a quantum system (a quantum
computer in the present context) and its environment. 
A pure state to be used as a computational resource
becomes a dirty mixed state due to decoherence and then
the computational outcome is not reliable any more.

There are several strategies to fight against
decoherence and quantum error correcting codes, 
abbreviated as QECC hereafter,
is one of the best weapons to fight against decoherence.
A pure state may be contaminated due to the
interaction between the system and the environment. Then one
may embed the quantum information to higher dimensional 
Hilbert space so that either (i) the error acted on
physical qubit may be identified by introducing 
the error syndrome measurement qubits so that the
initial quantum information is recovered after applying
appropriate corrections or (ii) the error
operator acts only on a part of the Hilbert space keeping
the initial quantum information intact. The second
QECC scheme is often called the ``error-avoiding''
coding due to this reason. 
Decoherence free subspace (DFS) and noiseless subsystem
(NS) are two popular examples of the second kind \cite{Zanardi1997,Zanardi1997a,Zanardi1998,Lidar1998,Kempe2001,Lidar2000,Lidar2001,Choi2006,Knill2000,Fortunato2003,Viola2001b,Lidar2013,Kondo2013,Byrd2006}.

In this paper, we consider the second approach to deal with
quantum channels in which all physical qubits involved
in coding suffer from the same error operators.
There are two relevant cases in which such error operators
are in action; (1) when the size of a quantum computer is
much smaller than the wavelength of the external disturbances
and (2) when photonic qubits are sent one by one through an optical fiber
with a fixed imperfection. In both cases, the qubits suffer from 
the same errors leading to decoherence. Another instance in
which such encoding is useful is that when Alice sends
quantum information to Bob (possibly billions light years away) without
knowing which basis vectors Bob employs. Then mismatching of the
basis vectors is common for all qubits and such mismatching is regarded
as collective noise.

In our previous publications, we reported the following results:
\begin{enumerate}
\item For a limited class of error operators $\{\sigma_x^{\otimes n}, \sigma_y^{\otimes n},
\sigma_z^{\otimes n}\}$, it is possible to iteratively implement encoding/decoding
circuits which protects $n-1$ logical qubits when $n$ is odd and
$n-2$ logical qubits when $n$ is even \cite{Li2011a}. 
When $n$ physical qubits protect $k$ logical qubits, the encoding rate is
defined by $k/n$. The asymptotic encoding rate obtained in \cite{Li2011a}
is $1$ as $n \gg 1$ for both cases.

\item For general error operators $W^{\otimes n}$, where $W \in {\rm SU(2)}$, 
we gave explicit recursive implementation of encoding/decoding circuits
for arbitrary number $n$ of physical qubits. We have shown that $n=2k
+1$ physical qubits protect $k$ logical qubits, leading to 
the asymptotic encoding rate of $1/2$ \cite{Li2011}. 

\item A qudit is a $d$-dimensional analogue of a qubit. It transforms
under the action of the fundamental representation of SU($d$).
(It should not be confused with a vector transforming under 
the action of a $d$-dimensional representation of
SU(2).) In \cite{Li2013}, we identified the subspace with the maximal 
dimension of the total Hilbert space of physical qudits when $d=2$ and
$3$, which is immune
to collective noise operators of the form $W^{\otimes n}$
where $W \in \mbox{SU($d$)}, (d=2,3)$. It was shown that the encoding
rate approaches to 1 as $n \gg 1$. The irreducible representation (abbreviated as irrep, hereafter) giving the encoding
subspace with the maximal dimension is given by an almost rectangular
Young tableau \cite{Li2013}. Identification of an irrep with the
maximal multiplicity for $d>3$ is a highly nontrivial open problem even
though the decomposition of $W^{\otimes n}$ into irreps is well
established. 
\end{enumerate}

In the present paper, we demonstrate why the recursion relation
introduced in \cite{Li2011} works from representation theory
point of view and generalize this relation to qudit case.
We show how to implement encoding/decoding
circuits for $n$ physical qudits, which results in the asymptotic
encoding rate of $1/d$.
A natural question to raise from this statement must be
``why do we do this analysis even though it is known that
there is a DFS/NS which gives asymptotic encoding rate of 1?''.
To implement encoding/decoding circuits, we need quantum
circuits, which {\it physically} represent the encoding/decoding
matrix $U_E/U_E^{\dagger}$. Although it may be possible to find the
quantum circuits for small $n$ by some trial and error, it is
totally impossible to find them if the number of qudits $n$
is more than 100 or even 10. We believe recursive implementation 
of the circuits is the only possible way to physically 
realize proposed scheme.

The rest of the paper is organized as follows. In the next section,
we outline the results of \cite{Li2011} for qubits from a representation theoretical  viewpoint
so that they can be easily generalized to the qudits cases.
In section 3, we give the detailed analysis of recursive implementation of
qudits encoding/decoding circuits and prove that this implementation
gives the asymptotic encoding rate of $1/d$. Section 4 is devoted to
summary and discussions.

\section{SU(2) recursion relation revisited}
\begin{figure}
\begin{tabular}{c}
\includegraphics{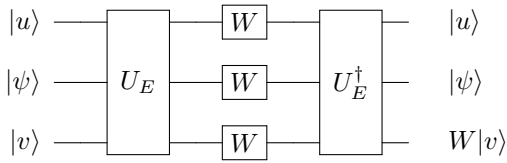}
\end{tabular}
\caption{Re-ordered version of the three qubit QECC from \cite{Li2011}. A circuit representation for $U_E$ is given is Fig. \ref{fig:qecc3-elementary-explicit}. $\ket{\psi}$ represents the data qubit. $\ket{v}$ is the state of the ancillary qubit, which can be arbitrary.}
\label{fig:qecc3-elementary}
\end{figure}

\begin{figure}
\begin{tabular}{c}
\includegraphics{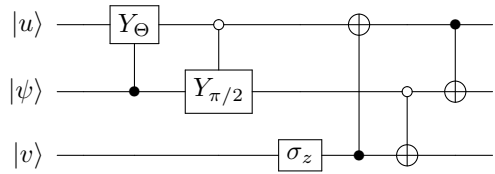}
\end{tabular}
\caption{Re-ordered SU(2) encoding gate $U_E$ from \cite{Li2011}, in terms of single-qubit and two qubit controlled-$U$ gates. Above $Y_\theta=\exp(i\sigma_y \theta)$ and $\sin\Theta=\sqrt{2/3}$.
}
\label{fig:qecc3-elementary-explicit}
\end{figure}

In this section, we review and give further explanation to the 3-qubit noiseless subsystem and recursion relation described in \cite{Li2011} from a representation theory point of view. This approach has the advantage of being general and applicable to systems with $d$ levels.

Let us denote the error acting on a single site as $W \in \text{SU}(2)$ and the total collective noise on the system as $\mathcal{E} = W \otimes W \otimes W$. Such an operation is totally symmetric under exchanges, and the resulting $8 \times 8$ matrix is reducible as $\mathbf 4 + \mathbf 2 + \mathbf 2$. In the context of representation theory, irreps of groups are conveniently labeled by Young tableau. The fundamental irrep of SU(2) is labeled as $\young(1)$. The form of the reduction is contained in the expansion \cite{Georgi1999}
\begin{eqnarray}
 \young(1) \otimes \young(1) \otimes \young(1) = \young(123) \oplus \young(12,3) \oplus \young(13,2).
 \label{eq:su2exp}
\end{eqnarray}
The irreps on the RHS have the dimensions of 4, 2, 2 respectively. The two copies of the fundamental irrep give rise to a noiseless subsystem. These irreps are more commonly known as spin-$3/2$ and spin-$1/2$ representations of SU(2) respectively. The dimension of an irrep is the number of vectors belonging to it whose entries are the Clebsch-Gordan coefficients, and they are sometimes called Young-Yamanouchi vectors \cite{Chen2002}.

If we denote the elements of the fundamental irrep as $u$ and $d$ (or $\ket{u}$, $\ket{d}$) which refer to the spin-up and spin-down states, the vectors belonging to the irreps that appear in Eq. (\ref{eq:su2exp}) can be written as
\begin{eqnarray}
\begin{aligned}
 \young(12,3) & \left\{ \begin{array}{l}
\begin{aligned}
 &  \frac{1}{\sqrt 6}\left(- [ud + du] u + 2 [uu] d \right) \\
 & \frac{1}{\sqrt 6}\left([ud + du] d - 2 [dd] u\right)
\end{aligned}
\end{array} \right. \\
\young(13,2) & \left\{ \begin{array}{l}
\begin{aligned}
& \frac{1}{\sqrt 2}(ud - du) u \\
& \frac{1}{\sqrt 2}(ud - du) d
\end{aligned}
\end{array} \right. \\
\young(123) & \left\{  \begin{array}{l}
\begin{aligned}
& uuu \\
& \frac{1}{\sqrt 3}(uud + udu + duu) \\
& \frac{1}{\sqrt 3}(ddu + dud + udd) \\
& ddd
\end{aligned}
\end{array} \right.
\end{aligned}
\end{eqnarray}
 The unitary transformation $U_E$ that block-diagonalizes $\mathcal E$ as $W \oplus W \oplus W^{(3/2)}$ where $W^{(3/2)}$ is the spin-$3/2$ representation of $W$ is constructed by using these basis vectors as columns and grouping them in a proper fashion such as \footnote{Here, the vertical dots indicate that vectors of the irrep are placed as column vectors.}
 \begin{eqnarray}
  U_E = \begin{pmatrix}
  \vdots & \vdots & \vdots \\
   \young(12,3) & \young(13,2) & \young(123)\\
   \vdots & \vdots & \vdots \\
  \end{pmatrix}.
\end{eqnarray}
An element of SU(2) is naturally expressed in an exponential form as $e^{i (r_x \sigma_x + r_y \sigma_y + r_z \sigma_z)}$.
Different representations can be obtained by replacing Pauli matrices, which correspond to the fundamental representation, with larger representations of the algebra $\mathfrak{su}(2)$. In the particular case of the 4-dimensional irrep, the generators are given as (see for example \cite{Pfeifer2003,Sakurai2010})
\begin{eqnarray}
\begin{aligned}
J_x^{(3/2)} &= \left(
\begin{array}{cccc}
 0 & \sqrt{3} & 0 & 0 \\
 \sqrt{3} & 0 & 2 & 0 \\
 0 & 2 & 0 & \sqrt{3} \\
 0 & 0 & \sqrt{3} & 0
\end{array}
\right), \\
 J_y^{(3/2)} &= i \left(
\begin{array}{cccc}
 0 & -\sqrt{3} & 0 & 0 \\
 \sqrt{3} & 0 & -2 & 0 \\
 0 & 2 & 0 & -\sqrt{3} \\
 0 & 0 & \sqrt{3} & 0
\end{array}
\right), \\
J_z^{(3/2)} &= \left(
\begin{array}{cccc}
 3 & 0 & 0 & 0 \\
 0 & 1 & 0 & 0 \\
 0 & 0 & -1 & 0 \\
 0 & 0 & 0 & -3
\end{array}
\right).
\end{aligned}
\end{eqnarray}

Figure \ref{fig:qecc3-elementary} shows the entire operation of sending the state $\ket{u} \ket{\psi} \ket{v}$ through noisy channel. Here, $\ket{u}$ is spin-up state, $\ket{\psi}$ represents the data and $\ket{v}$ in an arbitrary ancillary. This particular choice of the input state as well as the output given in this figure can be justified in the following manner. The density matrix of the system can be written as $(\ket{\psi}\bra{\psi} \otimes \ket{v}\bra{v}) \oplus 0_4$. The action of the reduced error operator $\mathcal E' = U_E^\dagger \mathcal E U_E = (\openone_2 \otimes W) \oplus W^{(3/2)}$ on this state only rotates $\ket{v}$, leaving $\ket{\psi}$ intact. Here $0_m$ is an $m \times m$ zero matrix and $\openone_m$ is an $m \times m$ identity matrix.
An important corollary is that the action of $\mathcal E'$ on the subspace $\ket{u} \ket{\psi} \ket{v}$ is equivalent to $\openone_2 \otimes \openone_2 \otimes W$. This enables recursive construction of noiseless subsystem for $2k+1$ qubits.

\begin{figure}
\begin{tabular}{c}
\includegraphics{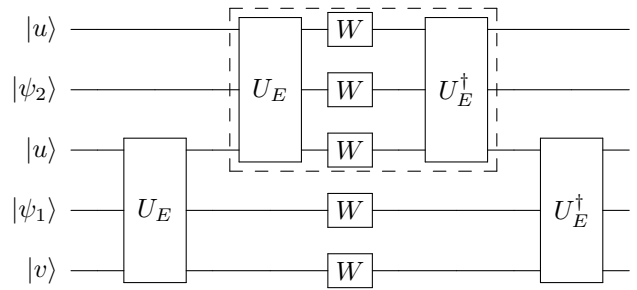}
\end{tabular}
\caption{Recursive 5-qubit circuit diagram from \cite{Li2011}. In this re-ordered version, the gates act on neighboring 3 qubits only.}
\label{fig:qecc3-5}
\end{figure}

\begin{figure}
\begin{tabular}{c}
\includegraphics{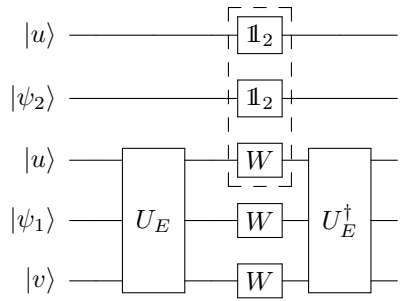}
\end{tabular}
\caption{Schematic recursion. This reduced circuit is equivalent to the one shown in Fig. \ref{fig:qecc3-5} due to the equivalence given in the text.}
\label{fig:qecc3-5-reduced}
\end{figure}

\begin{figure}
\begin{tabular}{c}
\includegraphics{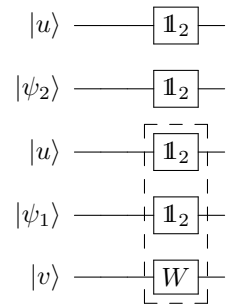}
\end{tabular}
\caption{The final version of the circuit given in Fig. \ref{fig:qecc3-5}.}
\label{fig:qecc3-5-reduced-2}
\end{figure}

For instance, to construct a noiseless subsystem for a 5-qubit system, we use $U_E$ twice as given in Fig. \ref{fig:qecc3-5}. By replacing the dashed part ($\mathcal E'$) of the circuit with $\openone_2 \otimes \openone_2 \otimes W$, we obtain the circuit given in Fig. \ref{fig:qecc3-5-reduced}. If we repeat the process for the lower 3-qubits, it becomes clear that the output is $\ket{u} \ket{\psi_2} \ket{u} \ket{\psi_1} (W\ket{v}$) and the states $\ket{\psi_i}$, $i=1,2$, are protected against noise (Fig . \ref{fig:qecc3-5-reduced-2}).

\section{Recursive Construction of Noiseless Subsystem for Qudits}
Now that we have the tools for a general analysis, we turn to the problem of finding an analogous recursion scheme for $d$-level systems.

To process, we first need to determine the number of qudits $m$ we require to avoid collective noise $\mathcal{E} = W^{\otimes m}$, where $W$ is an arbitrary error operator on a single qudit and an element of the fundamental representation of SU($d$). To construct a noiseless subsystem, we require to have an irrep with multiplicity that is at least $d$. It turns out that the fundamental irrep appears exactly $d$ times for $d+1$ qudits, which can be shown by using the Frobenius formula \cite{Chen2002}:
\begin{eqnarray}
(d+1)! \frac{\prod_{1 \leq i<j\leq d}\nu_i-\nu_j+j-i}{\prod_{i=1}^d (\nu_i+d-i)!} = d.
\end{eqnarray}
Here, $\nu_i$ denote the row lengths of the corresponding Young-diagram in a top-to-bottom order.
Such a noiseless subsystem can protect a single logical qudit against errors.

The collective error operator $\mathcal E$ is block-diagonalized by a unitary transformation $U_E^\dagger$ as
\begin{eqnarray}
 \mathcal E' = (\openone_d \otimes W) \oplus \mathcal O,
\end{eqnarray}
where $\mathcal O$ represents direct sum of the remaining representations of $W$, which are not relevant for our purposes.
The transformation matrix, which is the encoding circuit at the same time, is constructed by placing the Young-Yamanouchi vectors \footnote{The Young-Yamanouchi vectors are constructed from SU($d$) Clebsch-Gordan coefficients. Details on their computation can be found in \cite{Alex2011,Chen2002}} of the corresponding irreps as columns below
 \begin{eqnarray}
  U_E = \begin{pmatrix}
  \vdots & \vdots & \vdots & \vdots & \vdots \\
   \young(1) & \young(1) & \ldots & \young(1) & \text{other irreps} \\
   \vdots & \vdots & \vdots & \vdots & \vdots \\
  \end{pmatrix}.
  \label{eq:U_Egeneral}
\end{eqnarray}
Here each $\young(1)$ denotes any irrep that is equivalent to the $d$-dimensional fundamental irrep, and their ordering is not important and can be treated as a freedom during the construction of the encoding/decoding circuits.
In total, there are $d$ copies of this irrep. The vectors belonging to other irreps can be placed in an arbitrary manner. Note that in practice, we do not need to worry about these vectors as long as they are orthogonal to the basis vectors belonging to the fundamental irreps. Such an orthogonalization is enforced by the unitarity of the encoding circuit.

The proper input state turns out to be
\begin{eqnarray}
\ket{\Psi} = \ket{u}^{\otimes d-1}  \ket{\psi} \ket{v},
\end{eqnarray}
where $\ket{\psi}$ is the qudit state carrying the information and $\ket{v}$ is an ancillary qudit prepared in an arbitrary state and $\ket{u}$ is the $d$-dimensional vector $(1, 0, \ldots, 0)^T$, the highest-weight state in the fundamental representation of SU($d$).
Notice that the encoding and decoding can be seen as a basis transformation. In this view, $\ket{\Psi}$ has non-zero entries only in the first $d^2$ rows, which means such a state belongs to the direct-sum space of the fundamental irreps.

The action of the collective error $\mathcal E'$ on this state can be seen by acting on the corresponding density matrix $\rho = (\ket{\psi}\bra{\psi} \otimes \ket{v}\bra{v}) \oplus 0_{q}$ where $q=d^{d+1} - d^2$.
Clearly, $\ket{v}$ will be distorted into $W \ket{v}$ during the transmission through the noisy channel while the remaining qubits are left intact. We observe that the action of $\mathcal E'$ on this subspace is equivalent to $\openone_d^{\otimes d} \otimes W$, that is
\begin{eqnarray}
(U_E^\dagger W^{\otimes d+1}U_E) \ket{u}^{\otimes d-1}  \ket{\psi} \ket{v} = \ket{u}^{\otimes d-1} \ket{\psi} W\ket{v}
\label{eq:sud-gate}
\end{eqnarray}
holds. Following the arguments on Figs. \ref{fig:qecc3-5}, \ref{fig:qecc3-5-reduced} and \ref{fig:qecc3-5-reduced-2}, we see that the equivalence enables recursive construction of a $kd+1$-qudit QECC, which is capable of protecting $k$ qudits.

A naive way of constructing noiseless subsystem for $k$ qudits would be to vertically clone the elementary circuit such as the one given in Fig. \ref{fig:qecc3-elementary}. Since the elementary circuit protects a single qudit using $d+1$ qudits, the asymptotic encoding rate would be $1/(d+1)$.
However, with the recursive scheme, given that the number of correctable qudits using $n=kd+1$ for the channel is $k$, we find the asymptotic behavior of the encoding rate to be $k/n \to 1/d$ as $n \to \infty$ for a fixed $d$.

\section{Conclusion}
The noiseless subsystem is a method of using the inherent permutation symmetry of the noise to protect a subsystem against errors. In this work, we have used several powerful tools from representation theory for a better understanding and further generalization of the recursive construction of a subsystem for qubits, and extended our results to qudits. Our approach is based on a $d+1$-qudit encoding circuit whose implementation is realized by the vectors in the fundamental irrep $\young(1)$. It should be noted that different constructions based on different irreps are possible \cite{Byrd2006}, although they may not be necessarily suitable for our recursive scheme. We then generalized our construction to $n = kd + 1$. Encoding/decoding can be realized by using $U_E$/$U_E^\dagger$ successively, operating on $d+1$ neighboring qudits at a time, which can be of practical importance.

We note, however, that our construction does not give the maximum number of correctable qudits for the channel. When the irrep with maximal degeneracy is used instead of the fundamental representation, the ratio of protected qudits and total number of qudits is $k/n \to 1$ as $n \to \infty$ \cite{Li2013}. However, even though the DFS/NS with the maximal dimension is identified, we do not yet know how to implement the encoding circuit efficiently yet. Our study here gives a foolproof implementation of encoding circuit although the efficiency is $1/d$ for qudits. It is certainly desirable to find a recursion relation for maximal dimension DFS/NS, which is left as a future work.

It should be emphasized that the decomposition for $U_E$ given in Fig. \ref{fig:qecc3-elementary-explicit} is not canonical. In general, given a universal set of elementary gates, $U_E$ matrix can be decomposed in infinitely many different ways. Each decomposition has its trade-offs, some will require less energy or operational time than others for instance. Identification of ``good'' elementary gates (which are not necessarily 1- or 2-qudit gates \cite{Vartiainen2004}) and optimizing the decomposition in terms of these gates with respect to a cost function both require us to specify a Hamiltonian. Hence, both problems are implementation-dependent and no optimal generic decomposition exists. Once the Hamiltonian is decided upon, obtaining an optimal decomposition is still a non-trivial problem \cite{Vartiainen2005}.

Finally, we remark that our scheme is applicable to non-unitary error channels as well. The essential ingredient for our construction is the permutation symmetry of the collective error operator $\mathcal E$, and the Kraus operator $W$ may belong to a Lie group $G$ other than SU($d$) whose fundamental representation is $d$-dimensional, such as SL($d$, $\mathbb C$), following the Schur-Weyl duality. That is, the $U_E$ given in Eq. (\ref{eq:U_Egeneral}) will block-diagonalize $\mathcal E$ when $W \in \text{SL}(d, \mathbb C)$ and the resulting block structure will be the same \footnote{It should be noted that the reduced error operator $ \mathcal E'$ can be further reducible to irreps depending on $G$.}.

\section*{Acknowledgments}
We would like to thank Paolo Zanardi, Daniel Lidar and  Lorenza Viola for bringing some of the references to our attention.
UG and MN are grateful to JSPS (Japan Society for the Promotion of Science)
for partial support from Grant-in-Aid for Scientific Research (Grant Nos. 23540470 and 24320008).
UG acknowledges the financial support of the MEXT (Ministry of
Education, Culture, Sports, Science and Technology)
Scholarship for foreign students.
C.-K.L. was supported by a USA NSF grant, a HK RGC grant, and the 2011
Shanxi 100 Talent Program. He is an honorary professor of University of Hong Kong,
Taiyuan University of Technology, and Shanghai University. Y.-T.P. was
supported by a USA NSF grant and a HK RGC grant. N.-S.S. was supported
by a HK RGC grant PolyU 502512.


\bibliographystyle{apsrev}
\bibliography{extra,QECC}

\end{document}